\documentclass[%
 reprint,
 amsmath,amssymb,
 aps,
]{revtex4-1}

\usepackage{graphicx}
\usepackage{dcolumn}
\usepackage{bm}

\begin{document}


\title{Finite bias spectroscopy of a three-terminal graphene quantum dot in the multi-level regime}

\author{Arnhild Jacobsen}
\email{arnhildj@phys.ethz.ch}
\author{Pauline Simonet}
\author{Klaus Ensslin}
\author{Thomas Ihn}
\affiliation{Solid State Physics Laboratory, ETH Zurich, 8093 Zurich, Switzerland}


\date{\today}


\begin{abstract}
	
Finite bias spectroscopy measurements of a three-terminal graphene quantum dot are presented. Numerous lines of enhanced differential conductance are observed outside the Coulomb diamonds. In the single-level transport regime such lines are often associated with transport through excited states. Here the system is in the multi-level transport regime. We argue that the lines are most likely a result of strong coupling to only a few of the excited states available in the bias window. We also discuss the option that fluctuations of the density of states in the leads are fully or partly responsible for the appearance of the lines. Such a detailed analysis requires the presence of three leads to the dot. 

\end{abstract}



\maketitle



\section{Introduction}

Graphene quantum dots are considered as promising candidates for solid state spin qubits due to their predicted long spin lifetimes \cite{trauzettel2007}. In order to initialize, manipulate and read out such qubits, access to the discrete energy levels of the quantum dots are needed. Finite bias spectroscopy has been a convenient and widely used method to investigate the electronic structure of quantum dots in the past. During the last couple of years, also the direct measurement of excited states in  different graphene quantum dots including single layer single dots \cite{ponomarenko2008,schnez2009,guttinger2009,wang2011,volk2013}, single layer double dots \cite{molitor2010,liu2010} and bilayer double dots \cite{volk2011} have been reported. However, the lines of enhanced conductance parallel to the edges of the Coulomb diamonds due to transport through excited states are often accompanied by other lines of which the origin is not yet completely understood \cite{molitor2010,guttinger2009b}. Possible origins that have been suggested are modulation of the tunnel coupling due to resonances in the constrictions \cite{molitor2010,guttinger2009b} and phonon-mediated transport \cite{roulleau2011}.

Here we present finite bias spectroscopy measurement of a three-terminal graphene quantum dot. The main advantage of a three- compared to a two-terminal quantum dot is the possibility to get information about the individual tunnel barriers. In a standard transport experiment with a two-terminal quantum dot, the current through the dot is given by the average coupling of the dot wave function with the wave functions of the two leads \cite{leturcq2004}. Thus, it is not possible to investigate the two tunnel barriers separately. However, when three or more leads are connected to a quantum dot, the individual coupling strengths between the dot and the leads can be determined by measuring the complete conductance matrix of the system \cite{leturcq2004}.

Since graphene has no band gap, it is not possible to define nanostructures by  electrostatically confining the charge carriers. So far one common way to create tunnel barriers for graphene nano structures has been to etch narrow constrictions suppressing the current due to a disorder induced transport gap \cite{stampfer2008}. These tunnel barriers are complex systems themselves and have been studied extensively during the last couple of years \cite{han2007,todd2009,stampfer2009,droscher2011,bischoff2012}. In order to better understand transport through etched graphene quantum dots it is therefore useful to get separate information about the interplay between the dot wave function and the different lead wave functions. 

In a previous study of this three-terminal quantum dot the temperature dependence of Coulomb resonances and the width and shape of the resonances were thoroughly investigated and clear signatures of multi-level transport were found \cite{jacobsen2012}. The results presented here were performed during the same cool down and with identical experimental conditions as the previous study. In addition further measurements of the temperature dependence of Coulomb resonances strongly suggest that we are still in a regime of multilevel transport. Nevertheless, finite bias spectroscopy measurements of this quantum dot revealed a rich spectrum of lines of enhanced conductance outside the Coulomb diamonds. Based on the assumption of multi-level transport we discuss the possible origins of these lines by exploiting the additional information about transport through the individual leads obtained from the special three-terminal configuration.


\section{Device characterization}
\label{device char}


\begin{figure*}
  \begin{center}
    \includegraphics[width=1\textwidth]{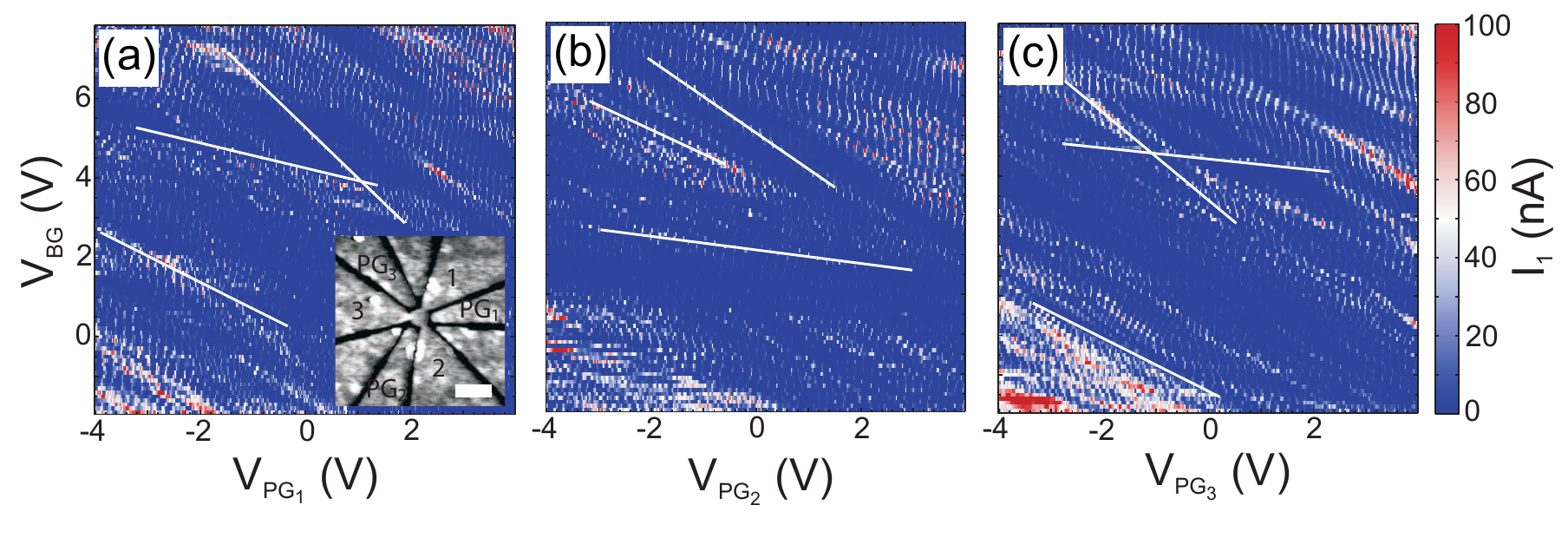}
	\caption{(a)-(c) The current in lead 1 as a function of back gate voltage and plunger gate voltage for PG$_1$, PG$_2$ and PG$_3$ respectively. The white lines represent the different relative lever arms. In the inset in (a) a scanning force microscopy image of the measured three-terminal quantum dot is shown. The three terminals are numbered from 1 to 3 and the three in-plane plunger-gates used to tune the device are marked with PG$_1$, PG$_2$ and PG$_3$. A bias voltage can be applied to each lead and the currents flowing through the different leads are measured. The scale bar corresponds to 200\,nm.
    }
    \label{fig1}
  \end{center}
\end{figure*}

The investigated device is made from a single layer graphene flake exfoliated from natural graphite and deposited onto a highly doped silicon substrate covered by 283\,nm of thermal silicon dioxide. In a first electron beam lithography (EBL) step followed by metal deposition (5\,nm Ti and 45\,nm Au) the ohmic contacts were defined. In a second EBL step followed by reactive ion etching (Ar and O) the quantum dot structure is patterned (for a detailed description of similar fabrication see Ref.\,\cite{guttinger2009b}).

A scanning force microscopy image of the final device can be seen in the inset in Fig.\,\ref{fig1}(a). The device consists of an island ($d=110$\,nm) connected to three leads (labelled 1, 2 and 3) by 40\,nm wide constrictions. Three different in-plane plunger gates labeled PG$_1$, PG$_2$ and PG$_3$ are used to tune the dot and the constrictions. A global silicon back gate is used to tune the overall Fermi energy of the device. The remaining three in-plane gates seen in the inset in Fig.\,\ref{fig1} (a) influence transport through the dot only weakly and are therefore not used.

We measure by applying a bias voltage to one lead while keeping the two other leads grounded. The currents flowing in all three leads are measured. All measurements presented in this study are carried out at a temperature of 1.7\,K.



To characterize the device we measure the current as a function of back gate voltage ($V_\mathrm{BG}$) and plunger gate voltage ($V_\mathrm{PG}$) in the Coulomb blockaded regime over a large range of gate voltages for all three plunger gates. This is depicted in Fig.\,\ref{fig1} (a)-(c) where $V_\mathrm{BG}$ is swept against $V_\mathrm{PG1}$, $V_\mathrm{PG2}$ and $V_\mathrm{PG3}$ respectively. In all three measurements a bias voltage of 1\,mV is applied to lead 1 and we plot the total current $I_1$ flowing through the dot.

In each of the three plots multiple diagonal lines characterized by three different slopes can be seen. These three slopes correspond to modulations of the current through the dot by localized states in the three constrictions \cite{stampfer2008}. Due to the large gate voltage ranges in these measurements the quantum dot conductance resonances cannot be seen. From the slopes we extract relative lever arms $\alpha_\mathrm{PG}/\alpha_\mathrm{BG}$ between all three plunger gates and the localized states in all three constrictions. A summary of these lever arms is shown in Table\,\ref{leverarms}. It can be seen that each plunger gate has a strong influence on localized states in the constriction directly to the right of the gate, which is also the constriction which is geometrically closest. The influence of a plunger gate on the constriction to its left is a bit weaker while the influence on the constriction on the opposite side of the dot is very weak. The slight asymmetry seen in comparable lever arms in Table\,\ref{leverarms} is due to the device being not perfectly symmetric and random resist residues on top of the device. For further characterization measurements see Ref.\,\cite{jacobsen2012}.

\begin{table}
	\begin{center}
    \caption{\label{leverarms}Relative lever arms $\alpha_\mathrm{PG}/\alpha_\mathrm{BG}$ for the in-plane gates with respect to the dot and the three constrictions.}
    \begin{tabular}{cccc}
    \hline
        & $\alpha_\mathrm{PG}^\mathrm{Constr. 1}/\alpha_\mathrm{BG}^\mathrm{Constr. 1}$ & $\alpha_\mathrm{PG}^\mathrm{Constr. 2}/\alpha_\mathrm{BG}^\mathrm{Constr. 2}$ & $\alpha_\mathrm{PG}^\mathrm{Constr. 3}/\alpha_\mathrm{BG}^\mathrm{Constr. 3}$\\
    \hline
        PG$_1$  & 1.15 & 0.68 & 0.25\\
        PG$_2$  & 0.13 & 0.88 & 0.65\\
        PG$_3$  & 0.65 & 0.13 & 1.15\\
    \hline
    \end{tabular}
	\end{center}
\end{table}



\section{Results and discussion}


\subsection{Coulomb Diamonds}

\begin{figure*}
  \begin{center}
    \includegraphics[width=1\textwidth]{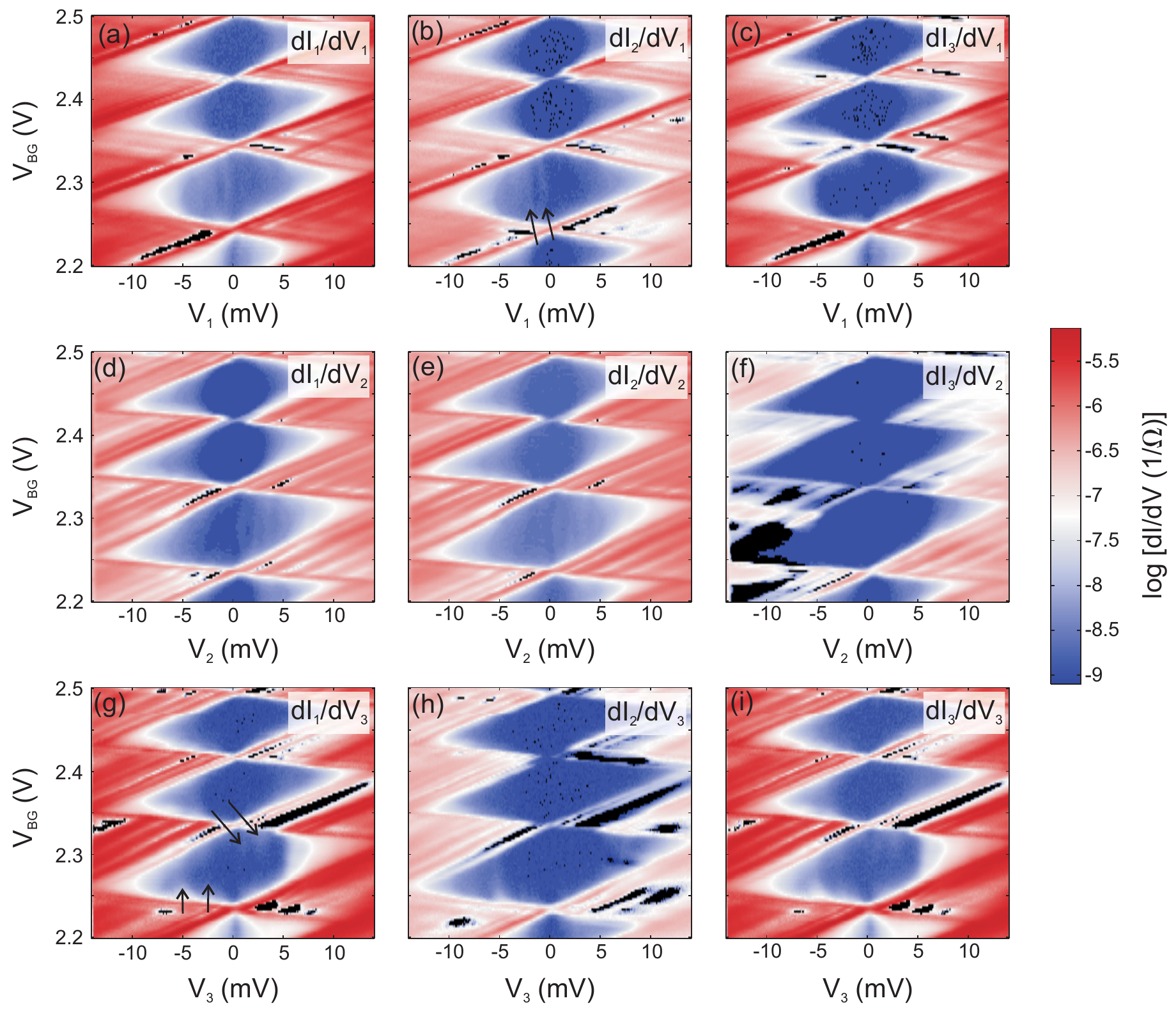}
    \caption{Differential conductance ($dI/dV$) as a function of bias voltage and back gate voltage. In the first row, (a)-(c), the bias voltage is applied to lead 1, in the second row, (d)-(f), the bias voltage is applied to lead 2 and in the third row, (g)-(i), the bias voltage is applied to lead 3. Correspondingly, the differential conductance measured in lead 1 can be seen in the first column ((a),(d),(g)), the differential conductance measured in lead 2 can be seen in the second column ((b),(e),(h)) and the differential conductance measured in lead 3 can be seen in the third column ((c),(f),(i)). Regions of negative differential conductance are marked in black.}
    \label{fig2}
  \end{center}
\end{figure*}

Fig.\,\ref{fig2} shows four consecutive Coulomb diamonds for which the complete matrix of the differential conductance was measured \cite{leturcq2004,jacobsen2012}. A bias voltage is applied to one of the three leads while the other two leads are grounded. In the first row the dc bias and the ac modulation voltage are applied to lead 1 [(a)-(c)], in the second row to lead 2 [(d)-(f)] and in the third row to lead 3 [(g)-(i)]. Similarly, the differential currents measured in lead 1, lead 2 and lead 3 are depicted in column 1, 2 and 3 respectively. As expected for a single confined dot, the diamonds are well defined and do not overlap. The general slight asymmetry of the diamonds is due to the asymmetric bias voltage. From these diamonds we determine the charging energy to be 15\,meV. This is the largest charging energy observed within this sample. Generally, the charging energies of most well defined Coulomb diamonds measured for this device are observed to fluctuate between 8 and 15\,meV. These charging energies agree well with charging energies published previously for devices of similar sizes \cite{schnez2009,guttinger2009,volk2013,guttinger2008}.

The most striking features seen in Fig.\,\ref{fig2} are the many lines of enhanced differential conductance outside the Coulomb blockaded regions that are parallel to the edges of the diamonds. Throughout the rest of this paper we will have a closer look at these lines and discuss their origin.


\subsection{Plunger gate dependence}

\begin{figure*}
  \begin{center}
    \includegraphics[width=1\textwidth]{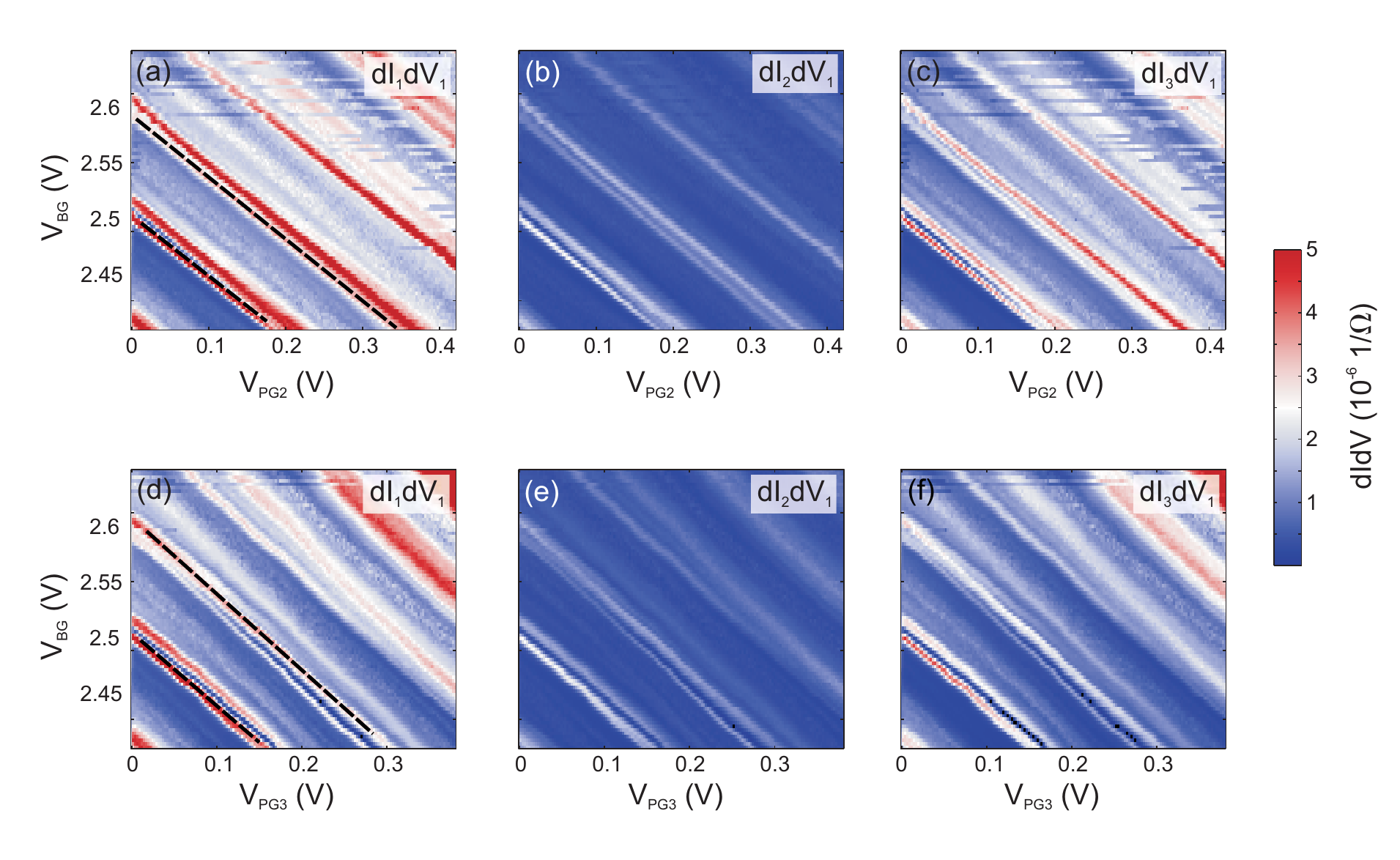}
    \caption{The evolution of the lines off enhanced differential conductance when sweeping the back gate against PG$_2$ ((a)-(c)) and PG$_3$((d)-(f)). A bias voltage of -12\,mV is applied to lead 1.
}
    \label{fig3}
  \end{center}
\end{figure*}

Lines of enhanced differential conductance outside Coulomb diamonds running parallel to the edges of the diamonds are often attributed to conductance through excited states and their appearance can thus be used to estimate the single-particle level spacing of the quantum dot states. However, such lines can also appear due to other effects. Both fluctuations in the density of states in the leads and single charge traps close to the quantum dot can give lines of enhanced differential conductance in the stability diagram of the quantum dot \cite{pierre2009}.

When characterizing the device we determined the relative lever arms between each of the three plunger gates and a localization in each of the three constrictions. Now we fix the bias voltage at -12\,mV and sweep the back gate and the different plunger gates in the same manner as shown in Fig.\,\ref{fig1} (a)-(c), but over a much smaller range. The result can be seen in Fig.\,\ref{fig3} (a)-(c) where the differential conductance in lead 1, 2 and 3 are plotted as a function of $V_\mathrm{BG}$ and $V_\mathrm{PG2}$. Fig.\,\ref{fig3} (d)-(f) shows the same measurement, however, PG$_3$ is swept instead of PG$_2$. In Fig.\,\ref{fig3} (a) and (d) two representative lines of enhanced differential conductance are marked with black dashed lines. From these lines relative plunger gate lever arms $\alpha_\mathrm{PG}/\alpha_\mathrm{BG}$ are determined to range from 0.50 to 0.52. This agrees perfectly with the relative dot lever arms determined from previous characterization measurement (see Ref.\,\cite{jacobsen2012}). From Table\,\ref{leverarms} the relative lever arms of the different plunger gates with respect to specific localized states in the constrictions are known. They are significantly different from the dot lever arms. Hence, if the lines of enhanced conductance seen outside the diamonds in Fig.\,\ref{fig2} were due to these localized states, lines with slopes corresponding to those seen in Fig.\,\ref{fig1} (a)-(c) should be visible in Fig.\,\ref{fig3}. This is not the case, and therefore we conclude that the lines of enhanced differential conductance outside the Coulomb diamonds do not originate from the localized states in the constrictions identified before.


\subsection{Transport through excited states?}

\begin{figure}
   \begin{center}
    \includegraphics[width=0.5\textwidth]{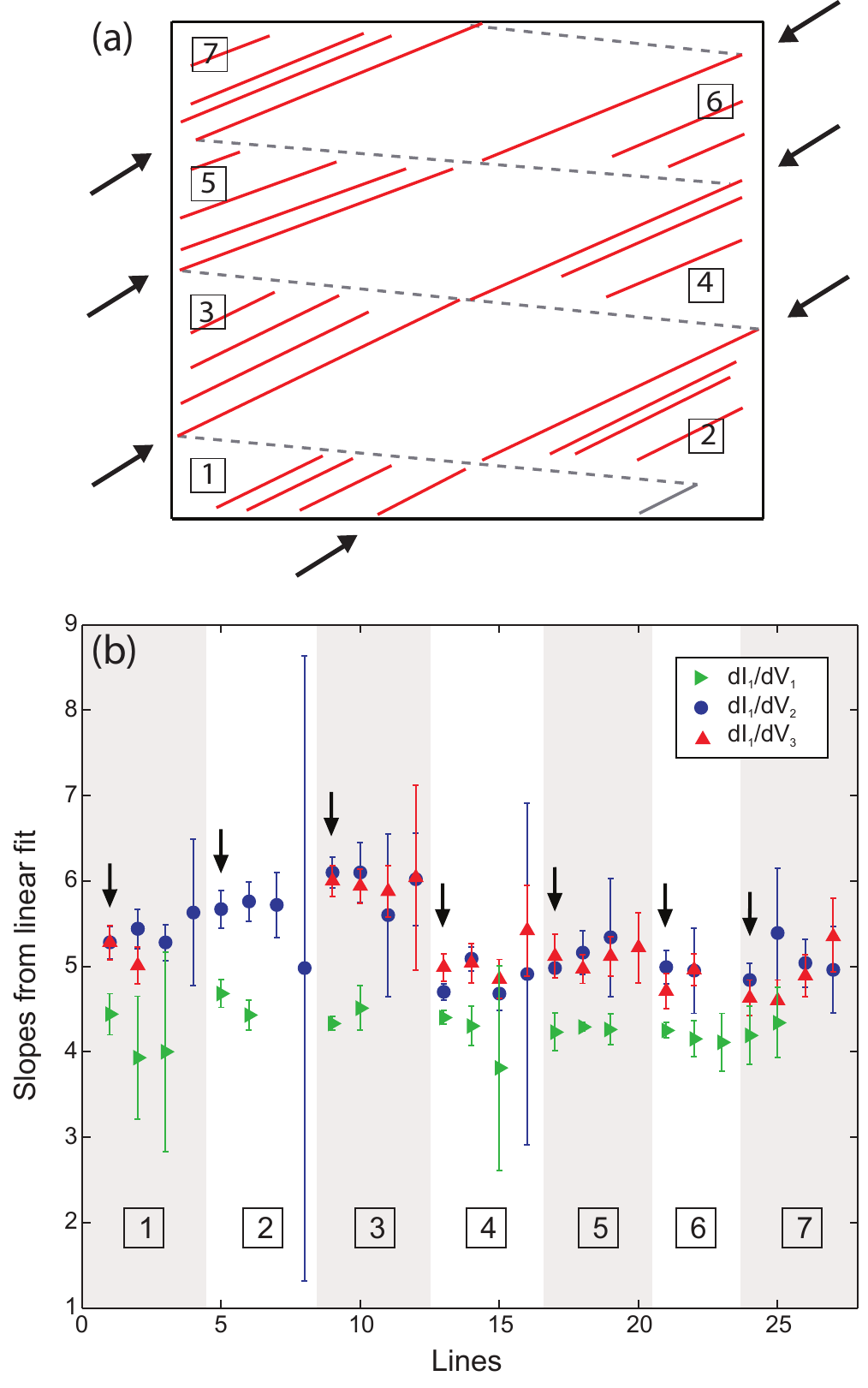}
\caption{Slopes of lines of enhanced differential conductance from Fig.\,\ref{fig2} are determined. As shown in the simple sketch of (a) we only consider lines parallel to the steeper diamond edge (where $\mu_N=\mu_S$) (red lines). In (b) the slopes determined from Fig.\,\ref{fig2} (a), (d) and (g) are plotted as green triangles, blue dots and red triangles respectively. Seven different regions, marked from 1 to 7 (see also (a)), have been evaluated. Grey regions correspond to slopes at negative bias voltages and white regions correspond to slopes at positive bias voltages. The first slope in each region, marked with an black arrow (see also (a)), is the slope of the diamond. The following slopes are from lines of enhanced differential conductance.}
    \label{fig4}
  \end{center}
\end{figure}

Next, we have a closer look at the diamonds depicted in Fig.\,\ref{fig2} and extract the accurate slopes of the different lines of enhanced differential conductance seen in this figure and compare them to the slopes of the diamond edges. As illustrated in the simple sketch in Fig.\,\ref{fig4} (a) we only consider lines parallel to the steeper diamond edge (where $\mu_N=\mu_S$ \cite{thijssen2008}) (red lines). In Fig.\,\ref{fig4} (b) all slopes for lines observed in Fig.\,\ref{fig2} (a), (d) and (g) are plotted. Since all plots measured in the same bias configuration [e.g (a)-(c)] have very similar lines of enhanced differential conductance, only one plot for each bias configuration is evaluated. Green triangles, blue dots and red triangles correspond to slopes extracted from Fig.\,\ref{fig2} (a), (d) and (g) respectively. Slopes from seven different regions, marked from 1 to 7 in the sketch in Fig.\,\ref{fig4} (a) have been extracted. Slopes within one of the grey shaded areas are extracted in a region of negative bias voltage while slopes within a white area are extracted in a region of positive bias voltage. The first slope in each region (marked with a black arrow) is the slope of the diamond edge itself. The following slopes within that region are the slopes of the lines of enhanced differential conductance, starting with the slope  of the line closest to the diamond edge. The slopes are extracted by making horizontal cuts of the 2D-maps, identifying the maxima and fitting them with a straight line. The error bars plotted are the 95$\%$ confidence interval of the slopes obtained from the linear fits. 

Looking at the slopes plotted in Fig.\,\ref{fig4} (b) no clear trends can be seen. For each of the three differential conductances only small, random fluctuations of the slopes can be seen. These fluctuations seem to be mainly caused by changes in the slopes of the diamond edges themselves and the lines of differential conductance only follow the trends of the diamond edges. Indeed, in 18 of the 20 analyzed regions the variation of the slopes within a specific region and for the same bias configuration is smaller than the error bars. Thus, within this analysis, there is no significant difference between the slopes of the diamond edges and the slopes of the corresponding lines of enhanced differential conductance. We therefore conclude that the lines of enhanced differential conductance outside the Coulomb blockaded regions are parallel to the diamonds edges within our measurement uncertainty. Thus, the lines could be really related to transport through excited states.

This is possible in a situation where the leads only couple strongly to a few of the excited states available within the bias window \cite{pfannkuche1995}. In such a case an excitation spectrum similar to that expected from single level transport will be observed. However, the level spacings will be larger than the true single-particle level spacing and might vary randomly. We find the spacing between the lines to vary randomly between 0.8\,meV and 3.5\,meV with most line spacings around 2\,meV. These values are not significantly larger than single-particle level spacings reported for a dot of similar size recently \cite{volk2013}. However, this might be due to a different number of charge carriers in the dot \cite{schnez2009}.

Previous measurements of the temperature dependence of Coulomb resonances showed behavior related to both single level transport and multi-level transport depending on which lead was probed \cite{jacobsen2012}. These observations, together with the observation of slight energy shifts of Coulomb resonances measured in different drain leads, were explained by a model where the different leads couple with different strength to the different dot-states in a regime of few-level transport. It is therefore likely that the lines of enhanced differential conductance seen here are a further result of this phenomenon.

Still, we can at this point not completely exclude that the lines are due to rapid changes of the density of states in the constrictions. It has been shown that the lines cannot be related to any of the localized states identified in the constrictions. However, on small voltage scales these localized states might change their geometry and furthermore the capacitances between them can change in such a way that it is possible to have local lever arms different than those found in Fig.\,\ref{fig1} (a)-(c). In addition, the existence of additional localized states, which could not be identified because of having lever arms too similar to those of the dot states cannot be excluded.

It should also be noted that we do not observe any clear inelastic co-tunneling onsets within the Coulomb diamonds. Such onsets are often used as an experimental proof of transport through excited states \cite{franceschi2001}. We do see similar features within the regions of suppressed current, marked with black arrows in Fig.\,\ref{fig2} (b) and (g), but these are oscillations and not the expected steps. The origins of these oscillations are so far not understood. Outside the Coulomb blockaded regions prominent regions of negative differential conductance (marked in black) associated with lines of enhanced differential conductance are observed. Such regions are regularly observed \cite{guttinger2009b} and may be due to both, transport through excited states, or lines due to modulations of the density of states of the leads.

Finally, in this study we have only considered transport through excited states and fluctuations of the density of states in the constrictions as possible origins for the lines of enhanced differential conductance outside the Coulomb diamonds. For suspended quantum dots in different material systems it has been shown that such lines can also appear due to phonon mediated transport \cite{weig2004,sapmaz2006,leturcq2009}. Here, we cannot completely exclude this possibility. However, the random and rather large spacings between the lines render this explanation unlikely \cite{guttinger2011}.


\section{Conclusion}

We have investigated four consecutive Coulomb diamonds where pronounced lines of enhanced differential conductance parallel to the diamond edges were observed outside the Coulomb blockaded regions. From a detailed analysis of the plunger gate dependencies and the slopes of these lines we conclude that they are most likely due to transport through a few of the available excited states which are strongly coupled to the leads. However, we also discuss the possibility that the lines originate from rapid fluctuations of the density of states of the constrictions.

The three-terminal setup allows us to carefully probe the occurrence of a specific transport feature in the current through a specific lead at zero and finite bias and measure relative lever arms to various gates. This combination, which is not possible in standard two-terminal configurations, is crucial for a detailed investigation and understanding of the excited states spectrum of graphene quantum dots. While our results do not allow to fully disclose the origin of the experimental features, they indicate the complications which need to be overcome before a quantitative understanding of the energy spectrum of graphene quantum dots can be obtained.	


\section*{Acknowledgements}
We thank D. Bischoff for helpful discussions.
Financial support from the Swiss Science Foundation is gratefully acknowledged.





\begin{thebibliography}{99}

\bibitem{trauzettel2007}
B. Trauzettel, D. V. Bulaev, D. Loss and G. Burkard
Nature Phys. \textbf{3}, 192 (2007)

\bibitem{ponomarenko2008}
L. Ponomarenko, F. Schedin, M. I. Katsnelson, R. Yang, E. W. Hill, K. S. Novoselov and A. K. Geim
Science \textbf{320}, 5874 (2008)

\bibitem{schnez2009}
S. Schnez, F. Molitor, C. Stampfer, J. G\"uttinger, I. Shorubalko, T. Ihn
and K. Ensslin
Appl. Phys. Lett. \textbf{94}, 012107 (2009)

\bibitem{guttinger2009}
J. G\"uttinger, C. Stampfer, F. Libisch, T. Frey, J. Burgd\"orfer, T. Ihn
and K. Ensslin
Phys. Rev. Lett. \textbf{103}, 046810 (2009)

\bibitem{wang2011}
L. -J. Wang, G. Cao, T. Tu, H. -O. Li, C. Zhou, X. -J. Hao, G. -C. Guo and G. -P. Guo
Chin. Phys. Lett \textbf{28}, 067301 (2011)

\bibitem{volk2013}
C. Volk, C. Neumann, S. Kazarski, S. Fringes, S. Engels, F. Haupt, A. M\"{u}ller and C. Stampfer
Nature Comm. \textbf{4}, 1753 (2013)

\bibitem{molitor2010}
F. Molitor, H. Knowles, S. Dr\"oscher, U. Gasser, T. Choi, P. Roulleau,
J. G\"uttinger, A. Jacobsen, C. Stampfer, K. Ensslin and T. Ihn
Europhys. Lett. \textbf{89}, 67005 (2010)

\bibitem{liu2010}
X. L. Liu, D. Hug and L. M. K. Vandersypen
Nanoletters \textbf{10}, 1623 (2010)

\bibitem{volk2011}
C. Volk, S. Fringes, B. Terr\'{e}s, J. Dauber, S. Engels, S. Trellenkamp and C. Stampfer
Nanoletters \textbf{11}, 3581 (2011)

\bibitem{guttinger2009b}
J. G\"{u}ttinger, C. Stampfer, T. Frey, T. Ihn and K. Ensslin
Phys. Status Solidi B \textbf{246}, 2553 (2009)

\bibitem{roulleau2011}
P. Roulleau, S. Baer, T. Choi, F. Molitor, J. G\"{u}ttinger, T. M\"{u}ller, S. Dr\"{o}scher, K. Ensslin and T. Ihn
Nature Comm. \textbf{2}, 239 (2011)

\bibitem{leturcq2004}
R. Leturcq, D. Graf, T. Ihn, K. Ensslin, D. D. Driscoll and A.C. Gossard
Europhys. Lett. \textbf{67}, 439 (2004)

\bibitem{stampfer2008}
C. Stampfer, E. Schurtenberger, F. Molitor, J. G\"uttinger, T. Ihn and K. Ensslin
2008 Nanoletters \textbf{8}, 2378 (2008)

\bibitem{han2007}
M. Y. Han, B. \"{O}zylmaz, Y. Zhang and P. Kim
%
Phys. Rev. Lett. \textbf{98}, 206805 (2007)

\bibitem{todd2009}
K. Todd, H. -T. Chou, S. Amasha and D. Goldhaber-Gordon
%
Nanoletters \textbf{9}, 416 (2009)

\bibitem{stampfer2009}
C. Stampfer, J. G\"{u}ttinger, S. Hellm\"uller, F. Molitor, K. Ensslin and T. Ihn
%
Phys. Rev. Lett. \textbf{102}, 056403 (2009)

\bibitem{droscher2011}
S. Dr\"{o}scher, H. Knowles, Y. Meir, K. Ensslin and T. Ihn
%
Phys. Rev. B \textbf{84}, 073405 (2007)

\bibitem{bischoff2012}
D. Bischoff, T. Kr\"{a}henmann, S. Dr\"oscher, M. A. Gruner, C. Barraud, T. Ihn and K. Ensslin
%
Appl. Phys. Lett. \textbf{101}, 203103 (2012)

\bibitem{jacobsen2012}
A. Jacobsen, P. Simonet, K. Ensslin and T. Ihn
New J. Phys. \textbf{14}, 023052 (2012)

\bibitem{guttinger2008}
J. G\"uttinger, C. Stampfer, S. Hellm\"uller, F. Molitor, T. Ihn and K. Ensslin
%
Appl. Phys. Lett. \textbf{93}, 212102 (2008)

\bibitem{pierre2009}
M. Pierre, M. Hofheinz, X. Jehl, M. Sanquer, G. Molas, M. Vinet and S. Deleonibus
Eur. Phys. J. B \textbf{70}, 475 (2009)


\bibitem{thijssen2008}
J. Thijssen and H. Van der Zant
Physica Status Solidi (B) \textbf{245}, 1455 (2008)

\bibitem{pfannkuche1995}
D. Pfannkuche and S. E. Ulloa
Phys. Rev. Lett. \textbf{74} 1995, (1995)

\bibitem{franceschi2001}
S. De Franceschi, S. Sasaki, J. M. Elzermann, W. G. van der Wiel, S. Tarucha and L. P. Kouwenhoven
Phys. Rev. Lett. \textbf{86}, 878 (2001)

\bibitem{weig2004}
E. M. Weig, R. H. Blick, T. Brandes, J. Kirschbaum, W. Wegscheider, M. Bichler and J. P. Kotthaus
Phys. Rev. Lett. \textbf{92}, 046804 (2004)

\bibitem{sapmaz2006}
S. Sapmaz, P. Jarillo-Herrero, Y. M. Blanter, C. Dekker, H. S. J. van der Zant
Phys. Rev. Lett. \textbf{96}, 026801 (2006)

\bibitem{leturcq2009}
R. Leturcq, C. Stampfer, K. Inderbitzin, L. Durrer, C. Hierold, E. Mariani, M. G. Schultz, F. von Oppen and K. Ensslin
Nature Phys. \textbf{5}, 327 (2009)

\bibitem{guttinger2011}
J. G\"uttinger
Ph.D Thesis ETH Z\"urich Switzerland (2011)	




\end{thebibliography}
\end{document}